\documentclass[12pt]{iopart}

\usepackage{graphicx}
\usepackage{cite}
\usepackage{float}

\begin{document}

\title{Investigation of the reaction $pp\to{p}K^{0}\pi^{+}\Lambda$ in search of the pentaquark}

\author{M Nekipelov$^{1,2}$,
  M B\"uscher$^1$,
  M Hartmann$^1$,
  I Keshelashvili$^{1,3}$,
  V Kleber$^4$,
  V Koptev$^2$,
  Y Maeda$^5$,
  S Mikirtychiants$^2$,
  R Schleichert$^1$,
  A Sibirtsev$^6$,
  H Str\"oher$^1$ and
  Yu Valdau$^{1,2}$
}

\address{ $^1$ Institut f\"ur Kernphysik, Forschungszentrum J\"ulich, 52425 J\"ulich, Germany }
\address{ $^2$ Petersburg Nuclear Physics Institute, 188300 Gatchina, Russia }
\address{ $^3$ High Energy Physics Institute, Tbilisi State University, 0186 Tbilisi, Georgia }
\address{ $^4$ Physikalisches Institut, Universit\"at Bonn, 53115 Bonn, Germany }
\address{ $^5$ Research Center for Nuclear Physics, Osaka University, Ibaraki, Osaka 567-0047, Japan }
\address{ $^6$ Helmholtz-Institut f\"ur Strahlen- und Kernphysik, Universit\"at Bonn, 53115 Bonn, Germany }

\ead{m.nekipelov@fz-juelich.de}

\begin{abstract}
  The reaction $pp\to{p}K^{0}\pi^{+}\Lambda$ has been studied with the ANKE
  spectrometer at COSY-J\"ulich at a beam momentum of 3.65$\,$GeV/c
  in order to search for a possible signal of the pentaquark $\Theta^+(1540)$,
  decaying into the $pK^0$ system. By detecting four charged particles in the final state
  ($\pi^+$, $\pi^-$ and two protons),
  the $K^0$ and the $\Lambda$ have been reconstructed to tag strangeness
  production. It has been found that the $\pi^+\Lambda$ missing-mass spectrum
  displays no significant signal expected from the $\Theta^+(1540)$ excitation.
  The total cross section for the reaction $pp\to{p}K^{0}\pi^{+}\Lambda$ has
  been deduced, as well as an upper limit for the $\Theta^+$ production 
  cross section. The intermediate $\Delta^{++}K^0\Lambda$ state seems
  to provide a significant contribution to the reaction.
\end{abstract}

%%%\submitto{JPG}
\maketitle

\section{Introduction}
The reactions $pp{\to}NK\pi\Lambda$ have been intensively used for hadron spectroscopy at
beam momenta above 5$\,$GeV/c, since they offer an excellent opportunity to search for 
baryon and hyperon resonances by studying $\pi{N}$ and $\pi\Lambda$ final states, 
respectively~\cite{alexander,alexander2,klein,klein2,dunwoo,holmgren,caso,bierman,firebaugh}.
It also allows one to investigate the $KN$ system, which attracted a lot of
attention due to its putative coupling to the pentaquark baryon $\Theta^+(1540)$~\cite{diakonov}. 
Despite a number of experimental indications for the existence of such a state, recent negative 
results, mostly obtained at higher energies (see e.g. \cite{battaglieri} and  references therein), have cast
serious doubt on the existence of the pentaquark $\Theta^{+}$. Its current status is reviewed in \cite{pdata}. 
There is no theoretical investigation that reconciles both the positive and the negative observations, 
although it has been shown in \cite{titov} that the production of $\Theta^{+}$ should be strongly suppressed at higher 
energies. Hadronic experiments at low energies are believed to be crucial to clarify the situation. 

The main features of the available data for $pp{\to}pK^0\pi^+\Lambda$
~\cite{alexander,alexander2,klein,klein2,dunwoo,holmgren,caso,bierman,firebaugh} are as follows: 
a) Structures in the $\pi^+\Lambda$ and $\pi^+{p}$ systems have been clearly identified as $\Sigma^+(1385)$ 
and $\Delta^{++}(1232)$ resonances, respectively, providing the dominant contributions to the reaction cross section. 
In the $\pi^+{K}$ system the $K^\ast(892)$ resonance has  been detected, although its contribution is rather small.
b) An enhancement in the $K\Lambda$ invariant mass spectrum at 1.7$\,$GeV, which is expected because of the 
excitation of the $P_{11}(1710)$ and $P_{13}(1720)$ baryon resonances~\cite{pdata}, has been observed~\cite{klein,firebaugh}. 
In \cite{alexander} it has been argued that  this enhancement results from a kinematical reflection due to $\Sigma^+(1385)$ production.
c) The $KN$ and $\Lambda{N}$ invariant mass spectra have not been investigated yet. All published data suffer 
from rather limited statistics and a moderate energy resolution. 

It is thus important to re-examine this reaction with new measurements with improved statistical accuracy
and high mass resolution. Such a dedicated experiment has been performed with the magnetic spectrometer 
ANKE~\cite{anke} at the COoler SYnchrotron COSY~\cite{maier} at the Research Centre J\"ulich.

\section{The experiment}
\begin{figure*}[htb]
  \centering
  \includegraphics[width=12cm]{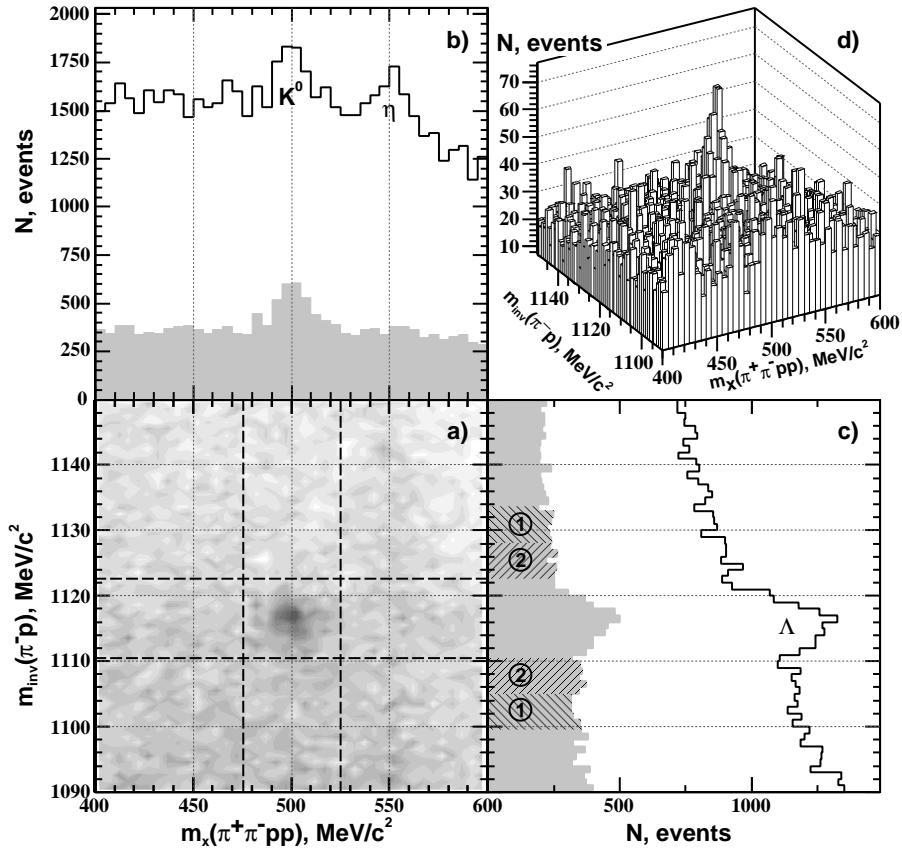}
  \caption{a) Mass contour plot: invariant mass of $\pi^{-}p$ versus missing 
    mass of $\pi^{+}\pi^{-}pp$. An enhancement corresponding to the masses of $K^{0}$ (497.7$\,$MeV/c$^2$) 
    and $\Lambda$ (1115.7$\,$MeV/c$^2$) is seen. Corresponding projections are presented in figures b and c.
    A lego plot is given in figure d. Shaded areas in the projection plots onto the axes $m_\mathrm{inv}(\pi^{-}p)$ (c) 
    and $m_\mathrm{x}(\pi^{+}\pi^{-}pp)$ (b) correspond to the independent cuts on masses of $K^{0}$ 
    and $\Lambda$, respectively. The dashed lines indicate these cuts. The hatched areas, 
    marked as 1 and 2 in the $m_\mathrm{inv}(\pi^{-}p)$ plot (c) denote the regions used for the background subtraction
    (see text).}
  \label{fig:2d}
\end{figure*}
COSY provided a circulating proton beam of $p_\mathrm{p}=3.65\,$GeV/c with an intensity of  approximately 
${4}\times{10}^{10}\,$protons per spill interacting with a hydrogen cluster jet target~\cite{cluster} with a thickness of
about  ${5}\times{10}^{14}\,\mathrm{{cm}^{-2}}$. The detection systems of ANKE, capable of detecting charged particles
with angles $\theta<12^\circ$, were triggered by a three-fold coincidence of two positively 
charged particles and a $\pi^-$ (from $\Lambda$ decay).
The reaction has been identified by the off-line selection of four particles, namely a $\pi^+$ and $\pi^-$ meson and two protons.
Detection and identification of all the charged particles was performed by a combination  of scintillation counters and multi-wire 
proportional chambers (MWPC). While the first allowed one to exploit the time-of-flight  technique, the MWPCs supplied 
full tracking and momentum information, and permitted the selection of particles originating from the target \emph{via}
analysis of their vertical angle (for details see ~\cite{anke,knim}). A coincidence measurement of four particles reduced 
the accidental background to a negligible level. The amount of misidentified particles was also small, due to the analysis of 
time-of-flight differences between detected particles. Therefore, the remaining background was almost entirely of physical origin.

\begin{figure*}[htb]
  \centering
  \includegraphics[width=12cm]{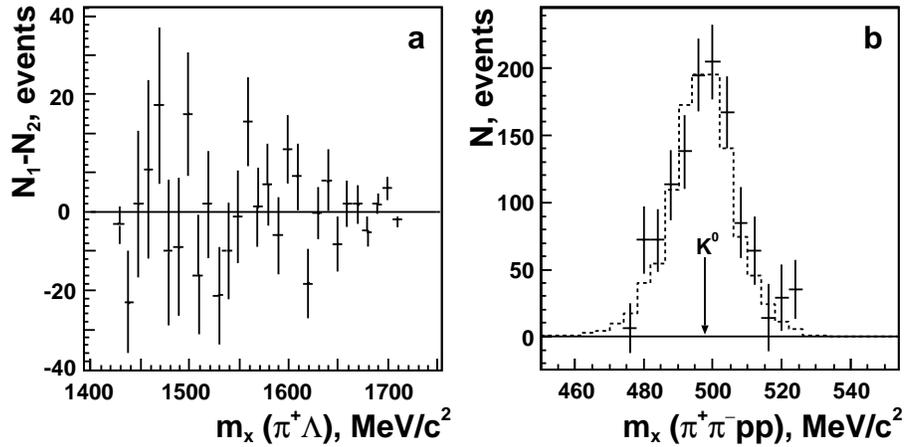}
  \caption{Side band subtraction technique. a) The difference of two missing mass spectra is plotted, in which 
    different background regions were selected for the subtraction. b) $K^0$ signal after background 
    subtraction has been applied. The dashed line shows the result of Monte Carlo simulations assuming a
    $pK^0\pi^+\Lambda$ phase space behaviour.}
  \label{fig:band}
\end{figure*}
The reaction $pp\to{p}K^{0}\pi^{+}\Lambda$ has been reconstructed by identifying the $\Lambda$ hyperon in the $\pi^-p$ 
invariant mass and the $K^0$ meson in the missing mass of the $\pi^+\pi^-pp$ system. The $K^0$ signal thus includes 
both contributions from $K^0_\mathrm{S}$ and $K^0_\mathrm{L}$. The corresponding invariant mass \emph{versus}
missing-mass scatter plot is shown in \fref{fig:2d}a. An enhancement of $K^0\Lambda$  coincidences is clearly 
visible on top of a sizeable background, remaining after cuts on masses are made  (see shaded areas in figures~\ref{fig:2d}b and c).
This background was removed by the side-band subtraction method. For the subtraction,  events in the regions besides the $\Lambda$ peak 
are selected, assuming that the background behaviour is smooth. The latter assumption was checked by 
choosing two different background regions (hatched areas marked as (1) and (2) in \fref{fig:2d}c), and building the bin-by-bin 
difference between them. The side-band subtraction method can be applied, if such a difference  spectrum does not exhibit 
any distinctive structures. For our case the result of this investigation is presented in \fref{fig:band}a for the 
missing mass distribution $m(\pi^{+}\Lambda)$, indicating that no signal shape distortion should occur because 
of background subtraction. The resulting $K^0$ missing mass spectrum after background subtraction is shown in \fref{fig:band}b.
The dashed line shows the result of Monte Carlo simulations assuming a $pK^0\pi^+\Lambda$ phase space behaviour,
which reproduces the experimental spectrum.

Differential distributions of the individual particles, such as momenta and scattering angles, are also well described 
by Monte Carlo simulations with the total reaction cross section as a free parameter and assuming four-body phase space 
distributions in the final state.
\begin{figure}[htb]
\centering
\includegraphics[height=6cm,angle=-90]{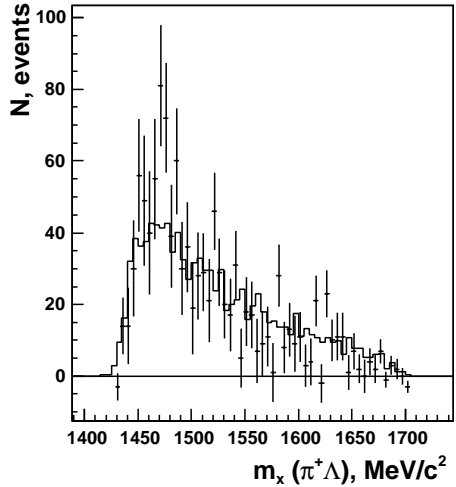}
\caption{Missing mass spectrum of $\pi^{+}\Lambda$ system. The solid line
  shows the result of Monte Carlo simulations assuming $pK^0\pi^+\Lambda$ phase space behaviour.}
\label{fig:penta}
\end{figure}
The missing mass distribution $m(\pi^{+}\Lambda)$, shown in \fref{fig:penta}, is the one  where the signal from the pentaquark
would be expected to appear. However, no obvious peak at 1.54$\,$GeV/c$^2$ is visible.
The total number of collected ${p}K^{0}\pi^{+}\Lambda$ events is 1041, which corresponds to a rate of 82$\,$events/day for a total 
integrated luminosity of 36$\,$pb$^{-1}$.  Together with the experimental data, results from the simulations are plotted as the  solid line in
\fref{fig:penta}. As one can see, only a moderate agreement ($\chi^2_{ndf}=1.41$) between experimental data and 
simulations based on four-body phase space is achieved. While for the high mass part of the spectrum the agreement 
is reasonable, an obvious excess of experimental events is observed around 1.47$\,$GeV/c$^2$. Such an enhancement 
may be connected with the excitation of some intermediate resonances.

Indeed, at higher energies the reaction $pp\to{p}K^{0}\pi^{+}\Lambda$ proceeds mainly through the production of intermediate resonances 
$\Delta^{++}(1232)$ and $\Sigma^+(1385)$~\cite{bierman, firebaugh}. It has also been argued that $N^*(1650)$ or $N^*(1710)$,
as well as an excitation of $K^*(892)$, might be important~\cite{firebaugh}.

In order to check these hypotheses a simultaneous fit of the two mass 
distributions, namely $m_\mathrm{x}(\pi^{+}\Lambda)$ on one hand and 
$m_\mathrm{inv}(\pi^{+}p)$ (for $\Delta^{++}(1232)$), or $m_\mathrm{inv}(\pi^{+}\Lambda)$ (for  $\Sigma^+(1385)$), or 
$m_\mathrm{x}(\pi^{+}p)$ (for $N^*(1710)$) on the other, has been performed. Since
the  beam momentum is  close to the $K^*(892)$ production threshold the relevant  mass region is not populated, and  therefore this
resonance does not contribute significantly. By introducing the $N^*(1710)$ resonance into the simulations 
the agreement with experimental data significantly decreases, thus unfavouring this channel.
\begin{figure}[htb]
\centering
\includegraphics[height=6cm]{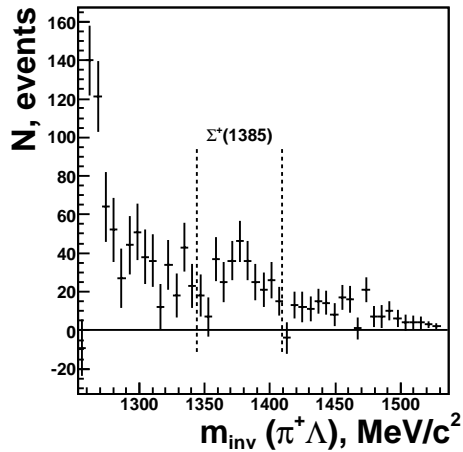}
\caption{a) Invariant mass spectrum of $\pi^{+}\Lambda$ system. 
The dashed lines restrict the area where a signal of $\Sigma(1385)$ is expected.}
\label{fig:sigma}
\end{figure}
No significant difference is found in the
quality of the fit including both the non-resonant channel and the one with an intermediate $\Sigma^+(1385)$, although
a small amount of events can be attributed to the production of  $\Sigma^+(1385)$ (see \fref{fig:sigma}). 
The fit results in a $\Sigma^+(1385)$ peak area, which is compatible with being a statistical fluctuation of the background. 
Therefore, only an upper limit for the possible $\Sigma^+(1385)$ production can be deduced.
Our estimates are based on the method of calculating the upper limit for the peak area as given in \cite{helene}.
This method assumes that the maximum peak area depends on the background and the total counts in the peak region.
The upper limit is then obtained for a given significance level, taking into account that the background is not exactly known and has a certain error.
As a result we extract $N_{\Sigma}<69\,$events at the $95\%$ confidence level, from which an upper limit for the cross section is deduced, given further below.

\begin{figure*}[htb]
\centering
\includegraphics[width=12cm]{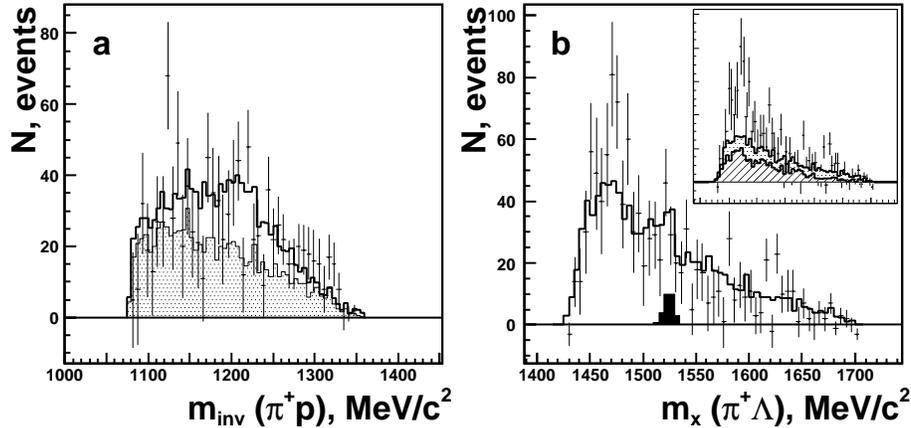}
\caption{a) Invariant mass of $\pi^{+}p$ system. The dotted area is obtained from the simulations assuming 
  phase space without intermediate resonances, and the solid line is the sum of all the contributions. 
  b) Missing mass spectrum of $\pi^{+}\Lambda$ system. The solid line denotes the sum
  of all the contributions. The inset shows individual contributions from non-resonant production (dotted) and from 
  intermediate $\Delta^{++}(1232)$ (hatched). The black region corresponds to the maximum permissible $\Theta^+(1540)$
  signal.
}
\label{fig:penta1}
\end{figure*}

The contribution from the formation of an intermediate $\Delta^{++}(1232)$ becomes obvious in the mass 
distribution of the $\pi^{+}p$ system shown in \fref{fig:penta1}a.  Here the dotted region originates from the 
$pK^0\pi^+\Lambda$ phase space, while the solid line depicts the sum of the latter and contributions from $\Delta^{++}(1232)$ 
and $\Theta^+$. The overall agreement between experimental data and simulations improves ($\chi^2_{ndf}=1.24$), as shown 
in \fref{fig:penta1}b. The inset in \fref{fig:penta1}b presents the individual contributions from non-resonant production 
and from intermediate $\Delta^{++}(1232)$ excitation. Since the shapes of the distributions above 1.47$\,$GeV/c$^2$ are 
very similar, any conclusion about the possible $\Theta^{+}$ production does not depend on the assumption 
made for ``non-exotic'' mechanisms.

Finally, the black area in (\fref{fig:penta1}b) corresponds to the possible signal expected from the $\Theta^{+}$ production. 
It has been found that the maximum permissible number of pentaquark events weakly depends both on the width and the position of the resonance. 
The width was varied from 1 up to 15$\,$MeV/c$^2$, with the latter value comparable to the missing mass resolution (FWHM) of the apparatus, while
the $\Theta^{+}$ mass ranged between 1.52$\,$GeV/c$^2$ and 1.54$\,$GeV/c$^2$.
The maximum number of pentaquark events found under these various assumptions is given by the best fit to the experimental data and equals to $28\pm{20}$.
It is obvious, that even with the best fit the signal is compatible with zero within errors, and therefore cannot be used for the cross
section estimates. Thus, as in the $\Sigma^+(1385)$ case, only an upper limit for the cross section will be given below.
For the evaluation of the upper limit of the $\Theta^+$ yield at the $95\%$ confidence level the approach 
from \cite{helene} has again been used. This resulted in 44 $\Theta^+$ events, used for calculation of an upper limit 
of the cross section (see below).

\begin{figure}[htb]
\centering
\includegraphics[height=6cm]{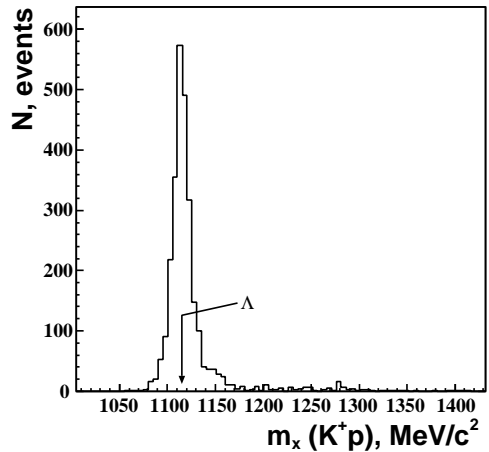}
\caption{Missing mass spectrum of $K^{+}p$ system in the reaction
  $pp\to{p}K^{+}\Lambda$. A background-free final-state identification has been achieved.}
\label{fig:lambda}
\end{figure}

For normalisation of the data, the reaction $pp\to{p}K^{+}\Lambda$ has been used. This 
reaction was measured simultaneously during the experiment. Again all four particles in the  final state have been detected,
including the proton and $\pi^-$ from $\Lambda$ decay. Contrary to the $pp\to{p}K^{0}\pi^+\Lambda$ reaction there are no background channels in the case 
of the production of ${p}K^{+}\Lambda$ (see \fref{fig:lambda}). Such a normalisation procedure, where a reaction with known cross 
section~\cite{lambda, sibir} and nearly the same final state is measured, allows one to avoid most of the systematic uncertainties connected with particle
identification and detector efficiencies, although adding a contribution due to the uncertainty in the acceptances of both measured channels.

\section{Results and discussion}
Since the acceptances of both four-body phase-space production and production with an intermediate
$\Delta^{++}(1232)$ are very similar, a total cross section for the $pK^0\pi^+\Lambda$ final state
can be calculated largely independent of the decomposition into separate channels. After the normalisation and efficiency corrections,
the following total cross section for the $pK^0\pi^+\Lambda$ final state has been deduced:
\begin{eqnarray}
\sigma_\mathrm{tot}=1.41\pm{0.05}\pm{0.33}\,\mathrm{\mu{b}},
\nonumber
\end{eqnarray}
where the first error is statistical, while the second is systematic. The systematic uncertainty is mostly coming from the 
error of $\sigma_{pK^+\Lambda}$~\cite{lambda}.

The total cross sections for the non-resonant channel and the channel with the $\Delta^{++}(1232)$ excitation have also been evaluated:
\begin{eqnarray}
\nonumber
\sigma^{\,\,\mathrm{non-resonant}}_{pK^0\pi^+\Lambda}=0.92\pm{0.16}\pm{0.21}\,\mathrm{\mu{b}},\\
\nonumber
\sigma_{\Delta^{++}K^0\Lambda}=0.49\pm{0.14}\pm{0.11}\,\mathrm{\mu{b}}.
\end{eqnarray}
Unlike with the total cross section, where the calculation relies on the number of experimental events, the results for the individual channels 
include an additional error from the decomposition. The measured cross section for the $pp\to\Delta^{++}K^0\Lambda$ reaction 
is significantly lower than a model prediction, $\sigma\approx{6}\,\mathrm{\mu{b}}$~\cite{sibir}.
However, this model overestimates as well the data available at high energies.

\begin{table}[htb]
  \caption{\label{tab:s}Contributions of different channels to the cross section for $pp\to{p}K^0\pi^+\Lambda$ in per cent
    of the total cross section $\sigma_\mathrm{tot}$ (see also \fref{fig:nrg}). NR denotes non-resonant production.}
  \lineup      
  \begin{indented}
  \item[]
  \begin{tabular}{cccccccc}
    \br
    Ref. & $p_\mathrm{p}$ &  $\sigma_\mathrm{tot}$ & NR & $\Delta^{++}$ & $\Sigma^{\ast{+}}$ & $K^*$ & $N^*$ \\
    & GeV/c   & $\mathrm{\mu{b}}$ & \% & \% & \% & \% & \%  \\
    \mr
    \emph{This}         &           &         & & & & & \\
    \emph{work}         &   3.65 & 1.41  & 65               &          35      & \emph{seen}     &    --  &                 --  \\
    \cite{klein}            &  6.70  & 64    &              36 &                36 &                     18 &    10 & \emph{seen} \\
    \cite{firebaugh}     &   7.87  & 72.4 &              5.7 &             34.2 &                   29.1 & 15.5 &              15.6 \\
    \br
  \end{tabular}
  \end{indented}
\end{table}
\begin{figure}[htb]
  \centering
  \includegraphics[width=6cm]{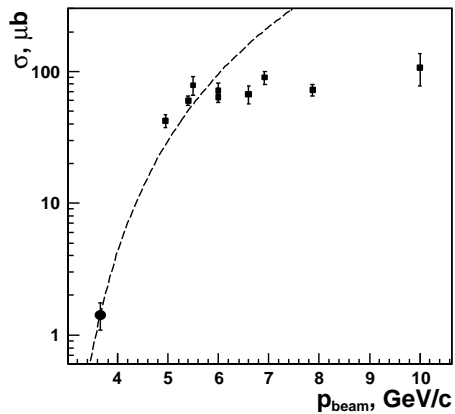}
  \caption{Total cross sections for the reaction $pp\to{p}K^{0}\pi^{+}\Lambda$ as a function of
    the beam momentum. The ANKE data point is denoted by a circle, while the other cross sections (see     
    \cite{alexander,alexander2,klein,klein2,dunwoo,holmgren,caso,bierman,firebaugh}) are given by squares.
    The dashed line corresponds to four-body phase space normalised to the ANKE data point.}
  \label{fig:nrg}
\end{figure}
The energy dependence of the total cross section of the reaction $pp\to{p}K^{0}\pi^{+}\Lambda$ is shown in \fref{fig:nrg}.
The data point from the current work is denoted by the full circle, while cross sections from other experiments are shown as squares. 
A nearly constant cross section above $p_\mathrm{p}\approx{6}\,$GeV/c can be understood if the 
reaction proceeds via intermediate resonances. Otherwise, 
a steep rise of the total cross section due to four-body phase space should be expected, as is shown by the dashed line in \fref{fig:nrg}. 
Such a resonance dominance 
has been observed at higher beam energies~\cite{bierman, firebaugh}, where $\Delta^{++}(1232)$ and $\Sigma^+(1385)$ 
provide the dominant contribution to the yield. Nevertheless, below $p_\mathrm{p}\approx{6}\,$GeV/c, where the energy available for the production
of the resonances is limited, non-resonant production is still competitive, and its relative contribution should decrease with increasing energy. The opposite behaviour 
is expected for resonance channels. Their contribution should increase with the beam momentum as illustrated in \tref{tab:s}.

Taking into account the estimated maximum number of 44 events for the $\Theta^+$ from the discussion above, 
we have evaluated an upper limit for the cross section of
$\Theta^{+}$-production in $pp\to\Theta^+\pi^+\Lambda$ at $p_\mathrm{p}=3.65\,$GeV/c:
\begin{eqnarray}
\nonumber
\sigma_{\Theta^+\pi^+\Lambda}<0.058\,\mathrm{\mu{b}}\qquad(95\%\,CL).
\end{eqnarray}
It is not possible to directly compare this result with the cross section for $pp\to{p}K^0\Sigma^+$ measured at 
$p_\mathrm{p}=2.95\,$GeV/c by the COSY-TOF collaboration~\cite{tof}, indicating a $\Theta^+$-signal.
The exit channels are distinct and even the number of particles in the final state differs.
As a crude estimate, one may assume that production of an additional pion leads to the reduction of the cross section by an order of magnitude. 
However, a similar factor can be regained due to the production of a $\Lambda$ instead of a $\Sigma$. But a difference in beam momentum of $700\,$MeV/c makes 
such a comparison very rather simplistic and questionable.

The upper limit for the $\Sigma^+(1385)$ production cross section is also provided by our data:
\begin{eqnarray}
\nonumber
\sigma_{\Sigma^+(1385)K^0p}<0.15\,\mathrm{\mu{b}}\qquad(95\%\,CL).
\end{eqnarray}
As has already been mentioned above,  the estimates of these cross sections do not depend on how the corresponding mass 
distributions are decomposed, since the shapes of the separate contributions are very similar in regions where the peaks from  $\Theta^{+}$
and $\Sigma^+(1385)$ are expected.

In summary, we have presented measurements of the reaction $pp\to{p}K^{0}\pi^{+}\Lambda$ for $p_\mathrm{p}=3.65\,$GeV/c.
While at higher beam momenta the formation of this final state is dominated by the production of intermediate resonances, 
non-resonant production seems to prevail at COSY energies, although the intermediate ($\Delta^{++}K^0\Lambda$)--state provides an important contribution to the yield. 
Unlike at higher energies, the contribution of $\Sigma^+(1385)$ is small. We have found no obvious signal indicative of the $\Theta^+(1540)$. 
Its cross section in $pp\to\Theta^+\pi^+\Lambda$ reaction is limited to be less than 58$\,$nb.  

\ack
We wish to acknowledge the assistance we received from the COSY and ANKE crews when performing these measurements.
This work has partially been supported by BMBF, DFG, Russian Academy of Sciences, and COSY FFE.

\end{document}